\documentclass[aip,jcp,reprint,onecolumn]{revtex4-1}
\usepackage{amsmath}
\usepackage{graphicx}
\usepackage{cleveref}
\usepackage{color}
\usepackage{caption}
\usepackage{subcaption}

\newcommand{\Fatt}{F_{\text{att}}}

\newcommand{\DG}{\Delta G}

\newcommand{\etal}{\textit{et al}}
\newcommand{\vv}[1]{\ensuremath{\mathbf{#1}}}
\newcommand{\vr}{\ensuremath{\mathbf{r}}}
\newcommand{\aaa}{\ensuremath{\alpha }}
\newcommand{\bbb}{\ensuremath{\beta }}
\newcommand{\DGij}{\ensuremath{\DG_{ij}(\vr_i,\vr_j)}}

\newcommand{\Or}{\ensuremath{\Omega(\{\vv{r}_i\})}}
\newcommand{\OR}{\ensuremath{\Omega(\{\vv{R}_I\})}}

\newcommand{\na}{n_{\aaa}}
\newcommand{\nb}{n_{\bbb}}
\newcommand{\naabb}{n_{\aaa\beta}}

\newcommand{\bna}{\ensuremath{\overline{n}_{\aaa}}}

\begin{document}

\title{Mobile linkers on DNA-coated colloids: valency without patches}
\author{Stefano Angioletti-Uberti}
\email{sangiole@physik.hu-berlin.de}
\thanks{Corresponding author}
\affiliation{Department of Physics, Humboldt University of Berlin, Newtonstr.15, 12489 Berlin, Germany}
\author{Patrick Varilly}
\affiliation{Department of Chemistry, University of Cambridge, Lensfield Road, CB2 1EW Cambridge, UK}
\author{Bortolo M. Mognetti}
\affiliation{Center for Nonlinear Phenomena and Complex Systems, Universit\'{e} Libre de Bruxelles, Code Postal 231, Campus Plaine, B-1050 Brussels, Belgium}
\author{Daan Frenkel}
\affiliation{Department of Chemistry, University of Cambridge, Lensfield Road, CB2 1EW Cambridge, UK}

\begin{abstract}
Colloids coated with single-stranded DNA (ssDNA) can bind selectively to other colloids coated with complementary ssDNA.
The fact that DNA-coated colloids (DNACCs) can bind to specific partners opens the prospect of making colloidal `molecules'.
However, in order to design DNACC-based molecules, we must be able to control the valency of the colloids, i.e. the number of partners to which a given DNACC can bind.
One obvious, but not very simple approach is to decorate the  colloidal surface with patches of single-stranded DNA that selectively bind those on other colloids. 
Here we propose a design principle that exploits many-body effects to control the valency of otherwise isotropic colloids. 
Using a combination of theory and simulation, we show  that we can tune the valency of colloids coated with mobile ssDNA, simply by tuning the non-specific repulsion 
between the particles. Our simulations show that the resulting effective interactions lead to low-valency colloids self-assembling in peculiar open structures, 
very different from those observed in DNACCs with immobile DNA linkers.
\end{abstract}

\maketitle
During the past two decades there has been substantial progress  in the functionalization of 
colloidal particles with various ligand-receptor pairs such as complementary single-stranded DNA (ssDNA) sequences \cite{mirkin,alivisatos}.
ssDNA grafting makes it possible to control the specificity of inter-particle
interactions \cite{tune1,tune2,tune3}:  two grafted ssDNA sequences bearing 
complementary Watson-Crick sequences can hybridise to form a
double-stranded DNA  (dsDNA) bridge between two particles, thus generating an effective attraction.
In contrast, particles coated with non-complementary sequences do not attract.
Exploiting this mechanism to tune colloidal interactions, DNA functionalisation has enabled the design 
of a variety of self-assembling nano-particle lattices~\cite{crystal-mirkin,crystal-gang, crystal-mirkin2}, 
thus opening the way towards new functional materials \cite{crystal-gang2}.\\
%
However, at present our ability to design arbitrary structures is limited by the fact that it is not straightforward to control the coordination number (i.e. valency) in such colloidal structures.
For instance,  low-valency colloids can self-assemble into open structures \cite{sciortino} that do not form if inter-particle interactions are pairwise additive and isotropic. 
On the atomic scale, carbon can form diamonds, where atoms are 4-coordinated, because carbon atoms have a well-defined electronic valency. In contrast,
noble-gases interact through (nearly) pairwise additive interactions and only form dense structures, such as fcc and bcc.\\ 
If we wish colloidal particles to self-assemble into a diamond lattice, we need to control their valency. 
Colloidal diamond lattices are intensively 
studied because 
such crystals would facilitate production of 
photonics band gap materials~ \cite{photonic,metamaterials}. However, their direct self-assembly is currently hampered 
by the lack of simple ways to control colloidal valency.\\
Considerable progress has been made in the (multi-step) synthesis of colloids with a well-defined valency encoded through the 
careful positioning of ssDNA linkers in patches at specific positions
. 
Wang \etal \cite{pine} have shown that it is possible to produce colloids with patches in precise locations; DNA can  be grafted selectively onto these patches.
In this  Letter, we present calculations that indicate that it should be possible 
to enforce the valency of colloidal particles without ``statically'' encoding
it in their structure. Instead, many-body effects naturally arising in 
DNACCs  with mobile linkers can be exploited to this purpose. Moreover we show
that valency control can be tuned by changing the grafting density of inert strands, temperature 
or salt concentration.\\

\begin{figure}
\begin{center}
\includegraphics[width=0.9\columnwidth]{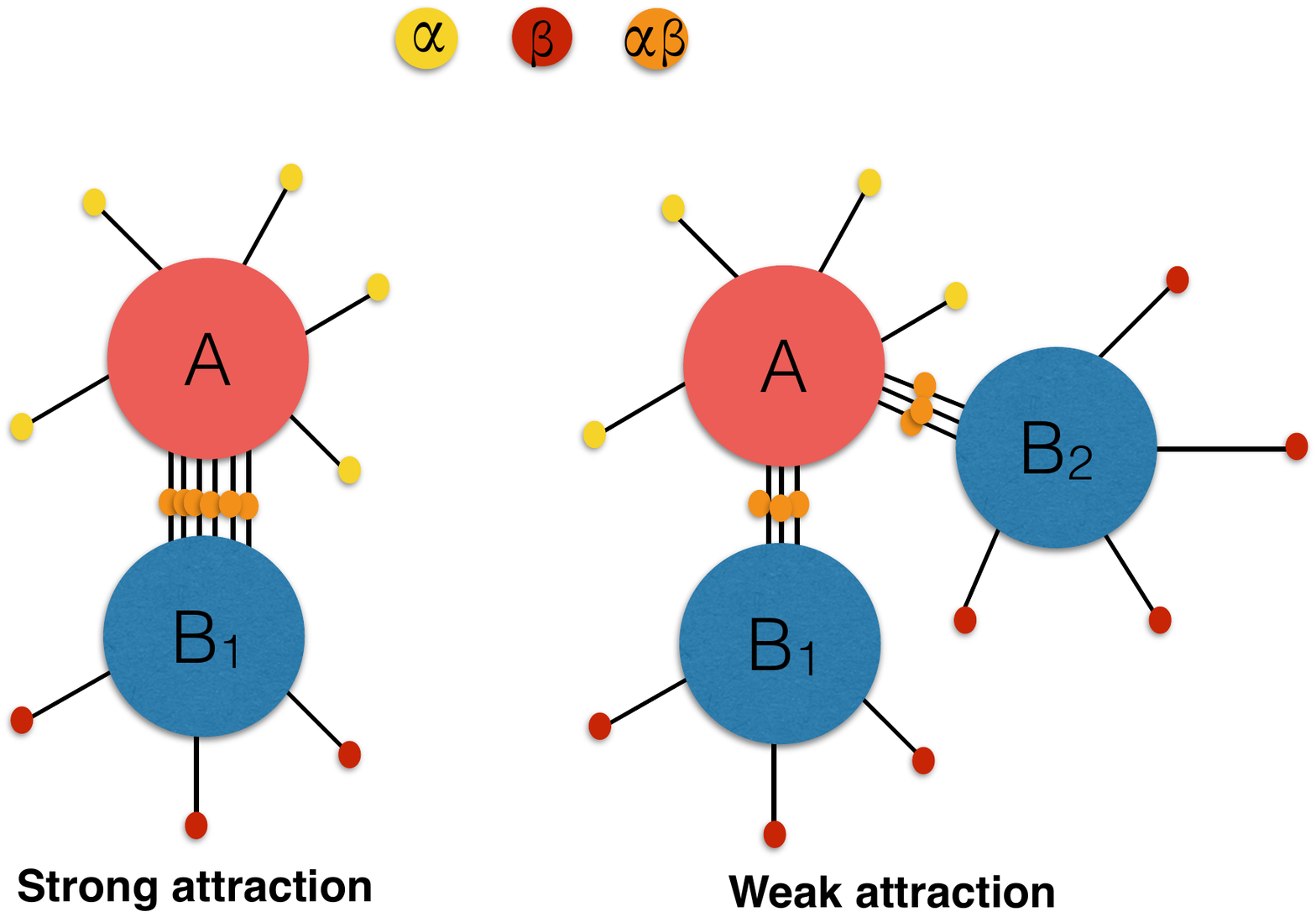}\\
\end{center}
\caption{
Schematic representation of our system. Colloid A bears DNA sequences that 
are complementary to those on B. Given that a fixed number of 
linking DNA exists, when multiple possible partners are present the total number 
of bonds between colloids decreases, hence their binding free-energy. 
This is the basis of the multi-body effect controlling valency.}
\label{fig:system}
\end{figure}

As an illustration we consider a binary system of colloids,  
$A$ and $B$ (see Fig.~\ref{fig:system}), covered with mobile $n_\aaa$ and $n_\bbb$ DNA strands. 
Each strand terminates in a short sequence of complementary ssDNA, \aaa~and \bbb.
Such colloids have been previously synthesised in various ways, as described in Refs.~\cite{mirjam-mobile,chaikin-mobile}.
When the suspension is cooled below a specific (sequence-dependent) temperature, the
ssDNA will hybridise with its complement, forming bonds between the DNACCs. 
Reliable techniques exist~\cite{patrick-jcp,stefano-jcp}  to predict the strength of attraction between  $A$ 
and $B$ colloids as a function of temperature. 
Same-type colloids (i.e. $A-A$ or $B-B$ pairs) repel each other 
due to the steric repulsion between non-binding ssDNA.\\
The interactions between colloids coated with mobile ssDNA are not pairwise additive. 
Consider  two DNACCs, $A$ and $B_1$, brought to a distance where hybridisation is possible. 
These two colloids will experience an attraction with a strength that increases with the number of bonds.
If a second colloid of type $B$  (here, $B_2$) is inserted in the system at 
the same distance from colloid $A$ as colloid $B_1$ (Fig.~\ref{fig:system}), any of the mobile DNA strands on $A$ can now hybridise with either $B_1$ or $B_2$.
The symmetry of the problem requires that on average the same number of bonds will form
between $A$ and $B_1$ and 
$A$ and $B_2$. Since the strength of the effective inter-particle interaction 
is an increasing function of the number of bonds 
and given that there is a finite number of strands to form bonds, 
the presence of a third colloid lowers the effective attraction between two particles. 
This many-body effect is at the basis of the mechanism controlling valency in this class of colloids. 
However, and this is our key point, the decrease of the binding strength per bond with the number of neighbours 
is {\em not} enough to control the colloidal valency, as
the maximum number of neighbours is determined by 
the total cluster interaction energy: each new bonding partner makes inter-particle interactions weaker 
but adds one more interacting pair. In the absence of non-specific repulsions, the highest coordination numbers are most favourable. 
However, if we add non-specific repulsions to the colloidal interactions, we can tune the optimal coordination number.\\
To make our argument quantitative, we calculate the effective interactions in different clusters.
To this end, we need an expression for the interaction free-energy of a cluster where the colloids positions are fixed at 
arbitrarypositions. We then relax the fixed-positions constraint and perform 
 Monte Carlo (MC) simulations where the colloids positions are allowed to 
achieve their equilibrium distribution.\\
Our expression for the effective interaction between DNACCs is based on the mean-field approach developed in 
Refs.~\cite{patrick-jcp,stefano-jcp}, and used to describe a variety of systems \cite{bortolo-pnas,stefano-nature,walkers,bortolo-sm}.
As shown in ref.~\cite{stefano-jcp}, this approach yields quantitative agreement with MC
simulations.\\ 
Ref.~\cite{stefano-jcp} showed that the attractive part of the 
effective interaction free-energy induced by a system of ligand-receptor pairs (e.g.
complementary DNA-strands) with bonding energies $\beta\Delta G_{ij}$ 
(where $i$ and $j$ label two specific binding partners) is
approximated remarkably well by the following expression:
\begin{equation}
\beta\Fatt = \sum_i \ln p_i + \sum_{i<j} p_{ij}
\label{eq:free-energy}
\end{equation}
where $p_i$ is the probability that linker~$i$ is unbound and $p_{ij}$ is
the probability that linkers $i$~and~$j$ form a bond. These quantities are 
given by solving the following set of equations:
\begin{align}
p_{ij} &= p_i p_j e^{-\beta\DG_{ij}},\label{eq:pij}\\
p_i &= 1 - \sum_j p_{ij}\label{eq:pi}
\end{align}
where $\DG_{ij}(\vr_i,\vr_j)$ is the free energy for the formation of a single bond between the $i-j$ pair. 
The latter can be rewritten in a more insightful form as \cite{patrick-jcp,mirjam-jcp}:
\begin{equation}
\beta\DGij = \beta\Delta G_0 + \beta\Delta G_{cnf}(\vr_i,\vr_j)
\label{eq:bond}
\end{equation}
where $\beta\Delta G_0$ is the hybridisation free-energy for two DNA strands in solution. 
$\beta\Delta G_0$ depends only on DNA sequence and is a function of temperature and salt concentration \cite{lorenzo-jacs, santalucia}. 
$\beta\Delta G_{cnf}(\vr_i,\vr_j)$, an explicit function of the grafting 
points $\vr_i,\vr_j$, is the configurational cost associated with the bond formation, and has been previously quantified 
both for single and double-stranded DNA \cite{patrick-jcp,lorenzo-jacs}.\\
%
%
%
%
%
For the case of mobile DNA, all strands with the same recognition sequence that reside on
the same colloid are equivalent since they cannot be distinguished by their grafting position. 
In this case, the correct procedure is to replace $e^{-\beta\DGij}$ by its average over all possible grafting points. 
Hence, the effective, single-bond strength between types \aaa~and \bbb~residing on colloids $A$ and $B$, respectively, will 
be given by
%
\begin{align}
\Xi_{\aaa\bbb}\left( \vv{R}_A,\vv{R}_B \right) &= <\exp\left( -\beta \DG_{\aaa\bbb} \right)>_{\mid \vv{R}_A,\vv{R}_B} \nonumber \\
    &= { \int_{S_A,S_B} \exp\left[-\beta\DG_{\aaa\bbb}\right]d\vv{r}_\aaa d\vv{r}_\bbb \over S_A S_B }
\label{eq:bond-average}
\end{align}
where the average is taken keeping the centre of colloid $A(B)$ at $\vv{R}_{A(B)}$ fixed and $S_{A(B)}$ is the  area of the colloid.
In Eq.~\ref{eq:bond-average} we use greek subscripts to label a strand type (rather than \textit{specific} strands as in Eqs.~\ref{eq:pi},\ref{eq:pij}).
We follow this convention from now on.
Using Eqs.~\ref{eq:pij}, \ref{eq:pi} to replace $p_{ij}$ in Eq.~\ref{eq:free-energy}, and considering that strands of the same type are equivalent 
and hence have the same value for $\Xi$, we obtain:

\begin{equation}
\begin{cases}
p_\aaa  + \sum\limits_{\gamma=1}^{N_{types}} n_\gamma p_\aaa p_\gamma \Xi_{\aaa\gamma}\left( \vv{R}_\aaa, \vv{R}_\gamma \right) = 1 \\

\cdots \\ 

p_{N_{types}}  + \sum\limits_{\gamma=1}^{N_{types}} n_\gamma p_{N_{types}} p_\gamma \Xi_{{N_{types}}\gamma}\left( \vv{R}_{N_{types}}, \vv{R}_\gamma \right) = 1. \\
\end{cases}
\label{eq:self-const-gen}
\end{equation}\\
and 

\begin{equation}
\beta F = \sum_{\gamma} n_{\gamma} \left[ \ln p_{\gamma} + 1/2 \left ( 1 - p_{\gamma} \right) \right].
\label{eq:free-energy-gen}
\end{equation}\\
Eq.~\ref{eq:self-const-gen} is a system of $N_{types}$ equations, one for each possible non-equivalent strand in the system: 
its solution is an explicit function of all colloidal positions $\{\vv{R}\}$.
Hence, if two strands cannot bind because they are grafted on distant colloids, $\Xi = 0$ and the sum over $\gamma$ in 
Eq.~\ref{eq:self-const-gen} effectively runs only on strand types on neighbouring colloids.\\
Eqns.~\ref{eq:self-const-gen}, \ref{eq:free-energy-gen} are key results of this paper. They allow to calculate 
the bond-mediated binding energy for any two generic objects interacting via mobile linkers. 
We show in the SI that for mobile linkers these formulas become exact in the limit of large numbers of linkers.\\

Let us first consider  clusters made of $1$ colloid of type 
$A$ surrounded by $N_B$ colloids of type $B$ at equivalent positions ( by symmetry only two types 
of strands are present, \aaa~and \bbb ) for which Eqs.~\ref{eq:self-const-gen},\ref{eq:free-energy-gen} become:

\begin{equation}
\begin{cases}
p_\aaa + N_B n_\bbb p_\aaa p_\bbb \Xi = 1 \\
p_\bbb + n_\aaa p_\aaa p_\bbb \Xi = 1 \\
\end{cases}
\label{eq:sis_pi}
\end{equation}

and 

\begin{equation}
\beta F_{clus}^{bond} = n_\aaa ( \ln p_\aaa + 1/2 - p_\aaa/2 ) + N_B n_\bbb ( \ln p_\bbb + 1 / 2 - p_\bbb/2 ).
\label{eq:en_clus}
\end{equation}
Eq.~\ref{eq:en_clus} (closed-form solution in SI) gives only the contribution due to bonds formation between ligands, and is purely
attractive. For typical DNACCs realisations, other terms due to van-der-Waals forces or electrostatic
interactions are negligible at the binding distance between colloids of a few nanometers imposed by the DNA length 
\cite{mirjam-jcp}. Hence, their effect can be safely disregarded. 
However, other terms due for example to the presence of inert DNA-strands or other polymers can still be relevant. 
These polymers act as steric stabilisers via excluded volume interactions, giving a repulsive energy of general form: 
\begin{equation}
\beta F_{rep} = -k_B T \ln\left( { \Or \over \Omega_{free} } \right)
\label{eq:en_rep}
\end{equation}
where \Or is the partition function counting all accessible states of the 
polymers given the positions of the colloids $\{\vv{R}\}$ 
and $\Omega_{free}$ is the same partition function when the colloids are at infinite separation.
As for $\beta\Delta G_{cnf}$, the contribution due to Eq.~\ref{eq:en_rep} can be calculated exactly for selected 
polymeric architectures or otherwise computed with MC simulations \cite{patrick-jcp}.\\
To illustrate the effect of non-specific repulsion, first consider the case that $F_{rep}$ between two colloids has a constant value
$F_{rep}^{min}$. The total energy of a $1 A+N_B B$ cluster then has an additional term
$N_B F_{rep}^{min}$. Added to Eq.~\ref{eq:en_clus}, we obtain a closed analytical 
expression for the free-energy of a cluster $F_{clus}\left( n_\aaa, n_\bbb, N_B, \Xi, F_{rep}^{min} \right)$. 
If we divide $F_{clus}$ by the number of neighbours, we obtain the total energy per 
bonding pair $F_{pair} = F_{clus}/N_B$ (Eq.~18-22 in the SI).\\
\begin{figure}
\begin{center}
\includegraphics[width=0.4\columnwidth]{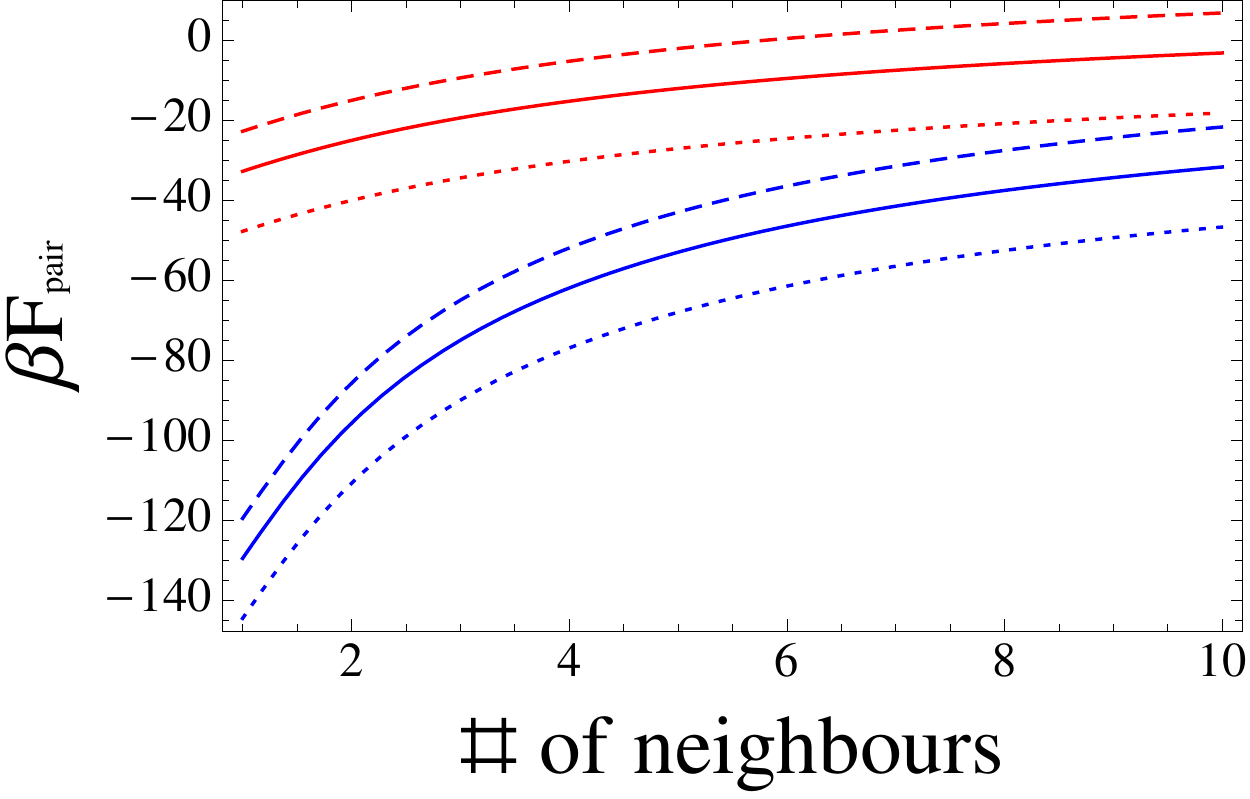}\label{fig:fpair}\\
\includegraphics[width=0.4\columnwidth]{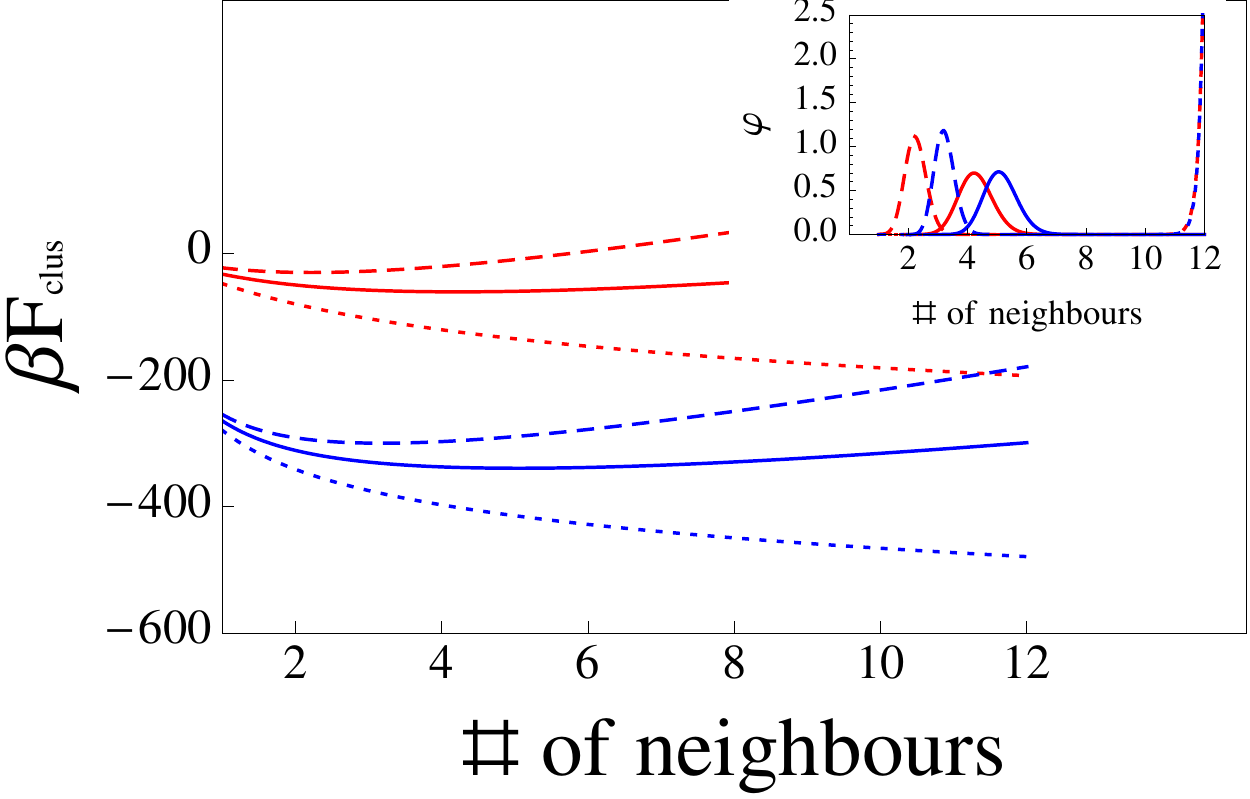} \label{fig:fpart}
\end{center}
\caption{ 
Energy per pair ($\beta F_{pair}$, above) and cluster energy ($\beta F_{clus}$, below)
for a cluster in our system of DNACCs with mobile strands. 
Same color (online version) means same value of $\Delta G_0$. Lines of different style represent 
different values of $\beta F_{rep}^{min}$ ($0$, $15$ and $25$ for dotted, continuous and dashed 
lines, respectively). Lines of different colours have been shifted by an irrelevant constant to facilitate 
comparison. The inset shows the valency probability distributions $\varphi$, which peak at low 
valency if $F^{min}_{rep}>0$.}
\label{fig:cluster}
\end{figure}
%
Fig.~\ref{fig:cluster} confirms that $F_{pair}$ is \textit{always} an increasing function of the number of colloids, 
hence attraction in an $A-B$ pair becomes weaker by increasing the number of neighbours. 
However, it is $F_{clus}$ that controls the valency distribution function. 
Without a local minimum in $F_{clus}$, the latter peaks at the highest possible coordination number (i.e. 12 for equal-sized spheres).
A minimum in $F_{clus}$ appears only if a finite repulsion $\beta F_{rep}^{min}$ is present, in which case the valency distribution 
peaks at a lower value dependent on $\beta F_{rep}^{min}$ (dashed and continuous curves in the inset), suggesting a viable
route to tune DNACCs' valency.

\begin{figure*}
\begin{center}
\includegraphics[width=0.45\columnwidth]{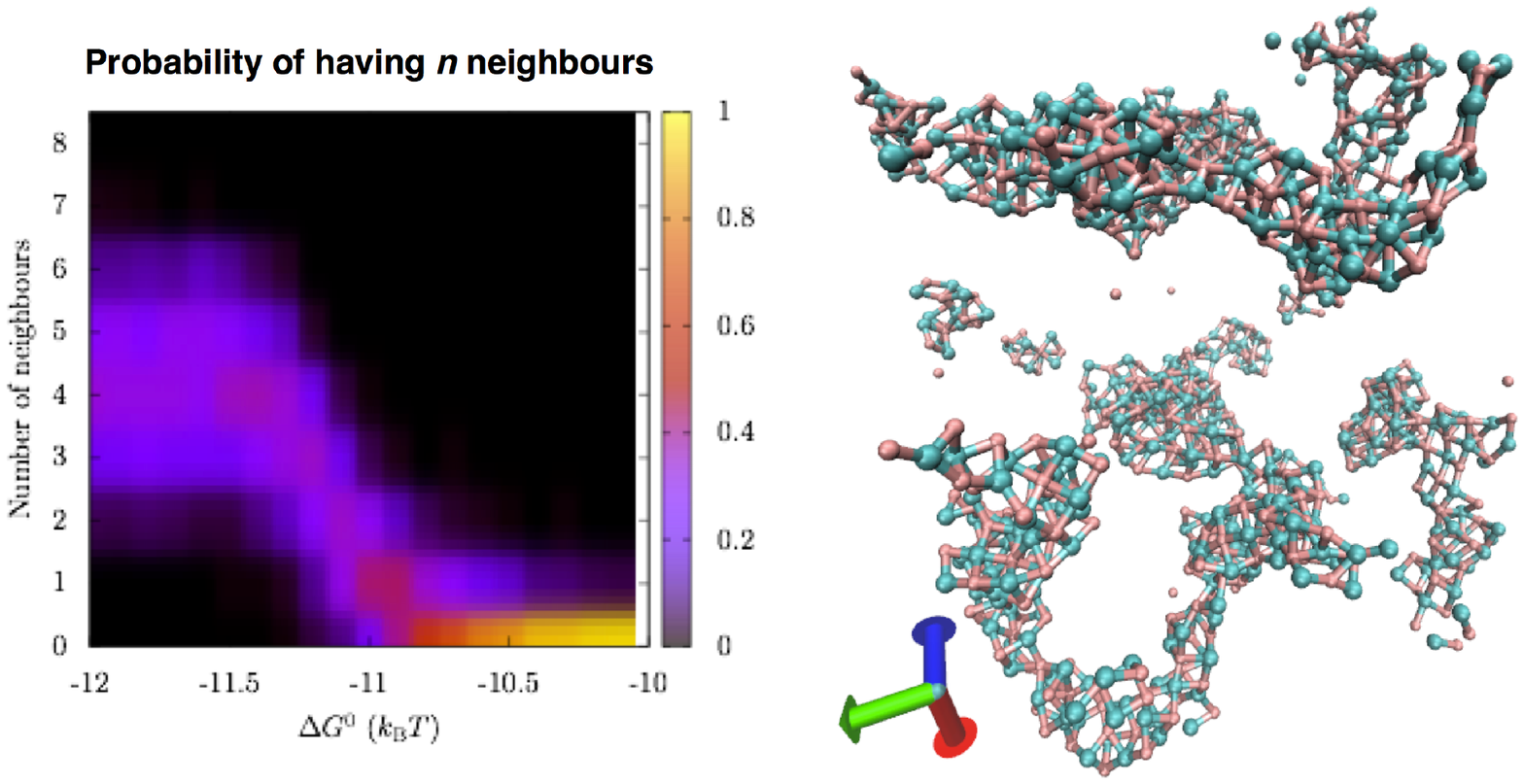}\includegraphics[width=0.45\columnwidth]{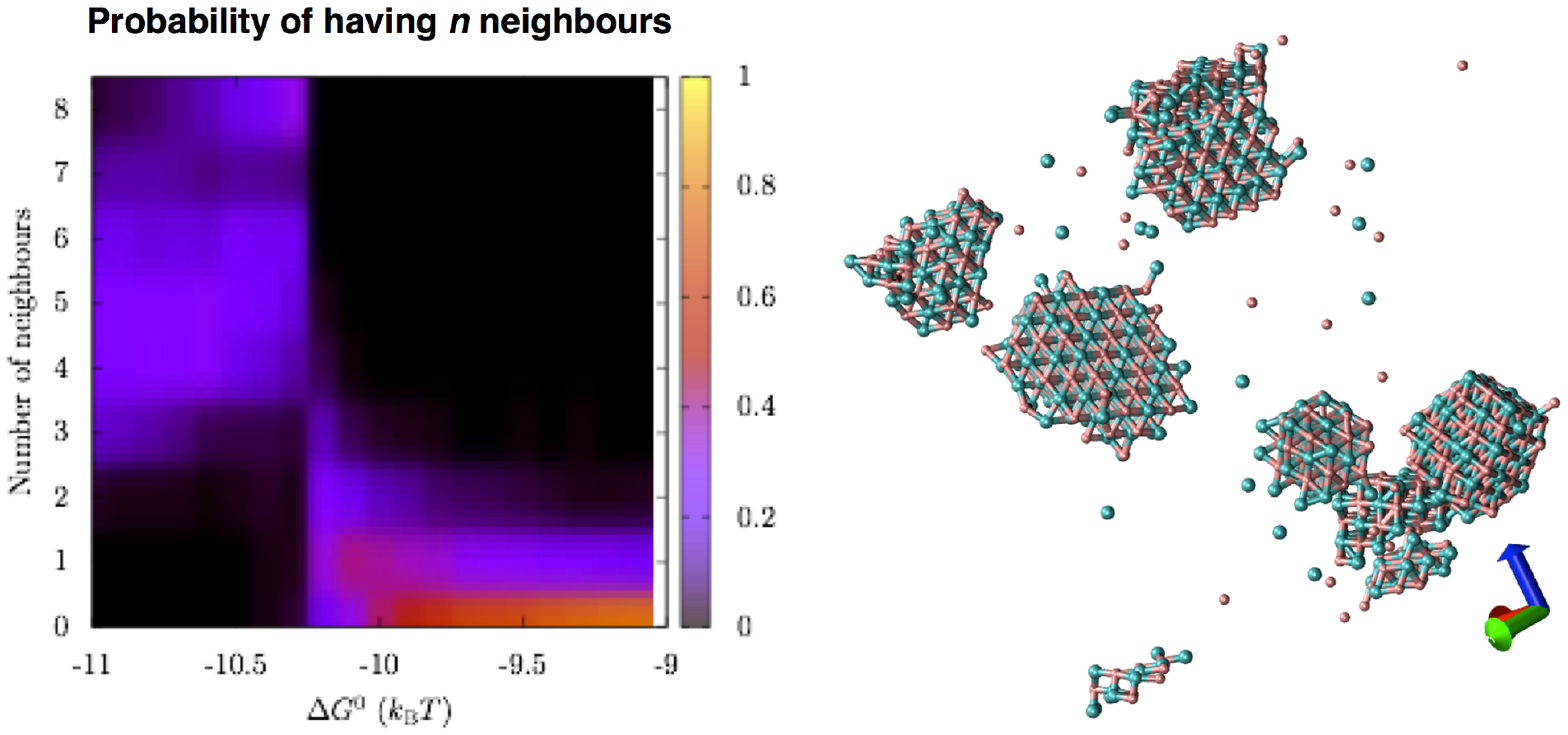}
\end{center}
\caption{Valency distribution as a function of $\beta\Delta G_0$ for colloids
with (left) and without (right) inert strands. The repulsive free-energy at the equilibrium distance
between colloids has an appreciable value only when inert strands are present, and is basically zero otherwise.    
The snapshots show typical configurations found at low $\beta\Delta G_0$ in the two cases, where the system either assembles
open structures of tetrahedral coordination (left, with inert strands) or a more compact NaCl structure (right, no inert strands).
}
\label{fig:snaps}
\end{figure*}

In practice, the repulsive energy at the equilibrium distance can be controlled 
by coating colloids with inert DNA strands or other polymers that are somewhat longer than 
the `sticky' DNA strands \cite{stefano-nature}.
Based on these results, we expect that in a realistic system of DNACCs
with mobile linkers one can control the average valency by varying temperature or salt concentration. 
We also expect, based on Eqs.~\ref{eq:bond-average},\ref{eq:en_rep}, that the specific value of 
$\Delta G_0$ at which a particular valency is stabilised will depend on the grafting 
density and the size of the colloids, since both these parameters enter in our equations.\\
To demonstrate this, we performed MC simulations of an equimolar $A:B$ mixture of colloids
that can move freely. 
$\beta F_{rep}$ was calculated by using Eq.~\ref{eq:en_rep} and considering the case of mobile strands 
(details of its calculation are reported in the SI). We stress that our outcomes are insensitive 
to the precise choice of $F_{rep}$.
%
We take two specific realisation of the system, differing in the presence or absence of long inert strands. 
Each colloid is modelled as a hard sphere with a radius $R=100$~nm on which $70$ rigid, double stranded DNA of length 
$L=20$~nm terminating with a short single-stranded DNA sequence are grafted (as in the plots for Fig.~\ref{fig:cluster}).
In the system with inert strands, $40$ additional strands of dsDNA of length $60$~nm are added.
Since $L<<\xi_p$, the persistence length of ds-DNA, linkers can be described as rigid rods \cite{mirjam-jcp}, 
for which both the contribution to the repulsive energy as well $DG^{conf}$ for mobile linkers 
can be calculated analytically given the colloids' positions (see the SI).
This model for the DNA-construct corresponds to the experimental realisation described in \cite{mirjam-nature,mirjam-self,melting-theory1,mirjam-jcp}.
%
In each run, $10^5$ MC sweeps {\em per} particles are made, starting with 100 colloids in random positions at 
packing fraction 0.05 and at various values of $\Delta G_0$. 
Each trial move consists of a random displacement $\vv{r}\in[-0.25 L, 0.25 L]^3$, and the total free-energy recalculated 
using Eqs.~\ref{eq:self-const-gen},\ref{eq:free-energy-gen} under periodic boundary conditions. 
%
The analysis was performed every 100 sweeps {\em per} particle, and the valency distribution function ($\varphi$ in the inset of Fig.~\ref{fig:cluster}) 
was calculated using the maximum bonding distance, i.e. $2L$ for rigid rods.\\
Results are presented in Fig.~\ref{fig:snaps}, for the case with (left) and without (right) inert strands, 
corresponding to $F_{rep}^{min}>0$ and $F_{rep}^{min}=0$, respectively.

%
%
%
These results support the conclusions based on the simpler analytical model
derived for the quenched-cluster system: repulsion plays an important role in stabilising low-valency structures. 
In particular, higher repulsion shifts the average valency to lower values.
As predicted by our simplified model, the valency probability distribution can be tuned 
by changing $\Delta G_0$, i.e. temperature 
or salt concentration.
The observed valency distribution for colloids without inert strands is relatively broad, 
which can mainly be attributed to finite size effects in our system.
Although we did not calculate the equilibrium phase diagram for this system, all observed structures assemble
quickly and spontaneously  from a random configuration and remain stable, suggesting 
at least metastability.
Without inert strands a compact and well ordered crystal forms, whereas the open structures observed in 
their presence lack long-range order. This is not necessarily required to achieve interesting functional properties:
low valency was shown to be enough to obtain structures with a proper, 3-dimensional photonic band-gap ~\cite{edagawa}.
\\
%
Finally, we note that Feng \textit{et al} ~\cite{chaikin-mobile} have reported the experimental observation of low-valency 
structures of deformable, micron-sized oil droplets coated with mobile DNA. In this system,  the repulsion mechanism is 
droplet deformation. As we have not applied our theory to this case, we cannot yet conclude whether droplet deformation 
alone can limit valency.
%

To conclude, in this this paper we have studied  the collective behaviour of 
a suspensions of binary \textit{isotropic} colloids functionalised by mobile linkers.
%
We showed how the interaction parameters 
can be tuned to induce 
the self--assembly of aggregates exhibiting a desired number of 
neighbours. Our model indicates that such a valency control can be achieved
by changing the non-specific repulsion between colloids and is a function of
temperature and salt concentration.
We also derive an explicit formula for the bonding-energy of a system of mobile linkers, provide the set of self-consistent
equations needed to calculate it, and show how they can be used to drive an MC algorithm to efficiently sample
the DNA-mediated free-energy. 
Hence beyond motivating experimental work towards the design of low
valency structures, we provide tools to model other 
systems
interacting via reversible mobile binders: an obvious example is the interaction between lipid vesicles \cite{herrmann}, or 
functionalised particles with cell membranes, whose interaction strongly depends on ligand-receptor bonds formation \cite{bell}.

\section{Acknowledgments}
S.A-U acknowledges support from the
Alexander von Humboldt Foundation via a Postdoctoral Fellowship.
This work was supported by the European Research Council Advanced
Grant 227758, the Wolfson Merit Award 2007/R3 of the Royal Society of London
and the Engineering and Physical Sciences Research Council Programme
Grant EP/I001352/1.  P.V. has been supported by a Marie Curie International
Incoming Fellowship of the European Community's Seventh Framework Programme
under contract number PIIF-GA-2011-300045, and by a Tizard Junior Research
Fellowship from Churchill College. 


%

\section{Supplementary Information}

\subsection{Accurate approximation for calculating the repulsive free-energy}

\begin{figure}
\centering
\begin{subfigure}[b]{0.48\textwidth}
\includegraphics[width=\textwidth]{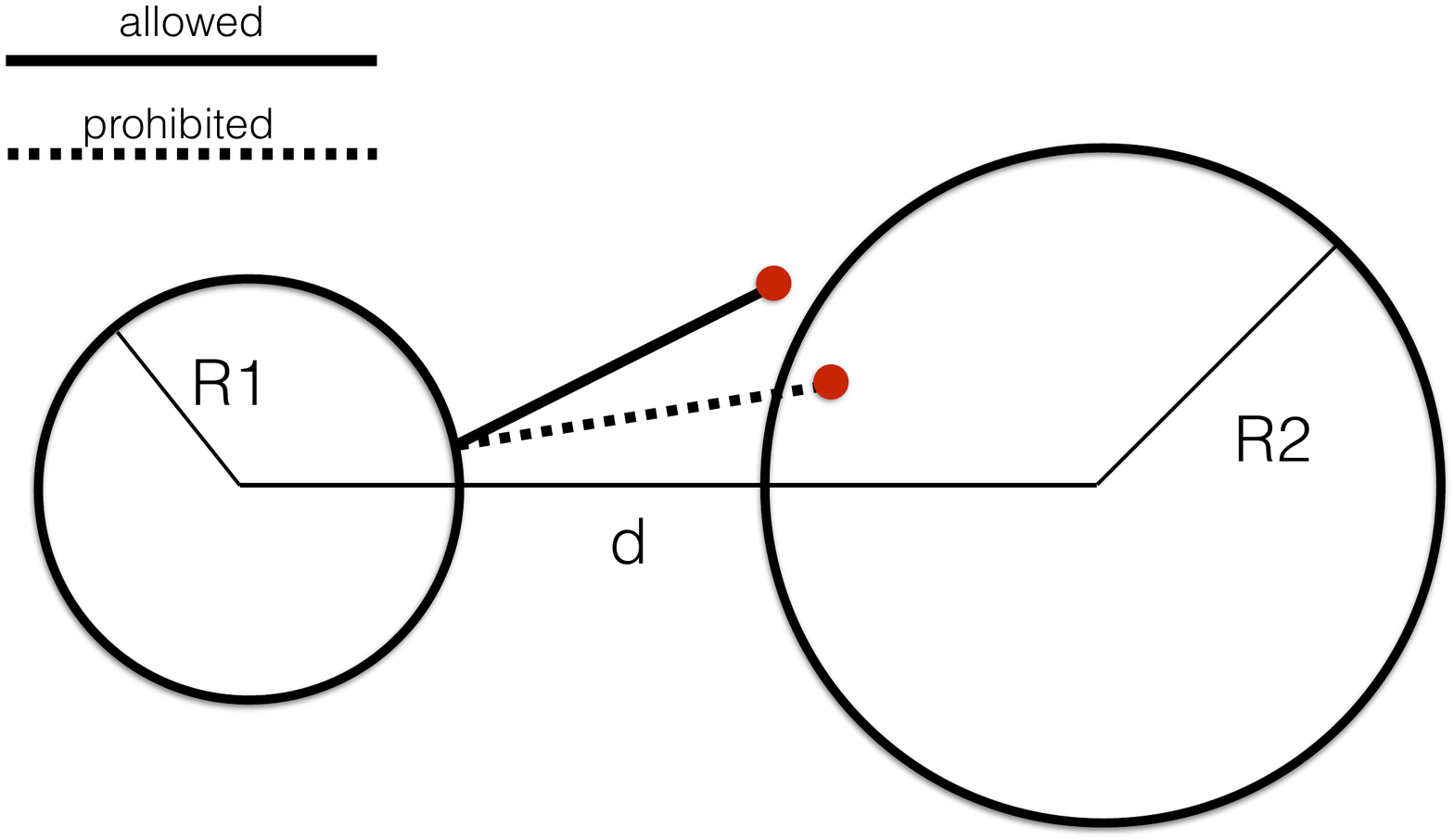} 
\caption{Allowed (continuous line) and prohibited (dashed line) state for a rod-like polymer grafted on the surface of a colloid.}
\label{fig:1a}
\end{subfigure}
\begin{subfigure}[b]{0.48\textwidth}
\includegraphics[width=\textwidth]{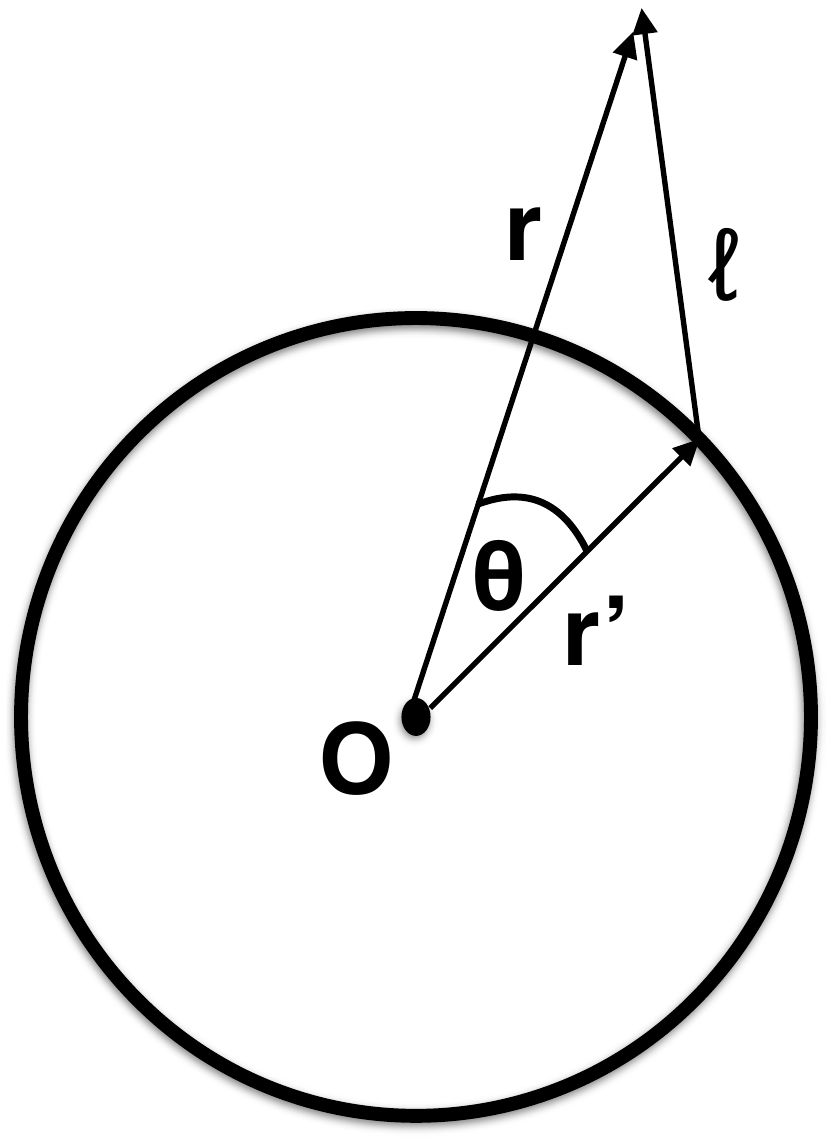} 
\caption{Geometry for the calculation of Eq.~\ref{eq:pr}. \vv{r}' is the vector giving the grafting position of the rod, whereas \vv{r} is the position of its endpoint. The rod length is fixed and equal to $\ell$.}
\label{fig:1b}
\end{subfigure}
\begin{subfigure}[b]{0.48\textwidth}
\includegraphics[width=\textwidth]{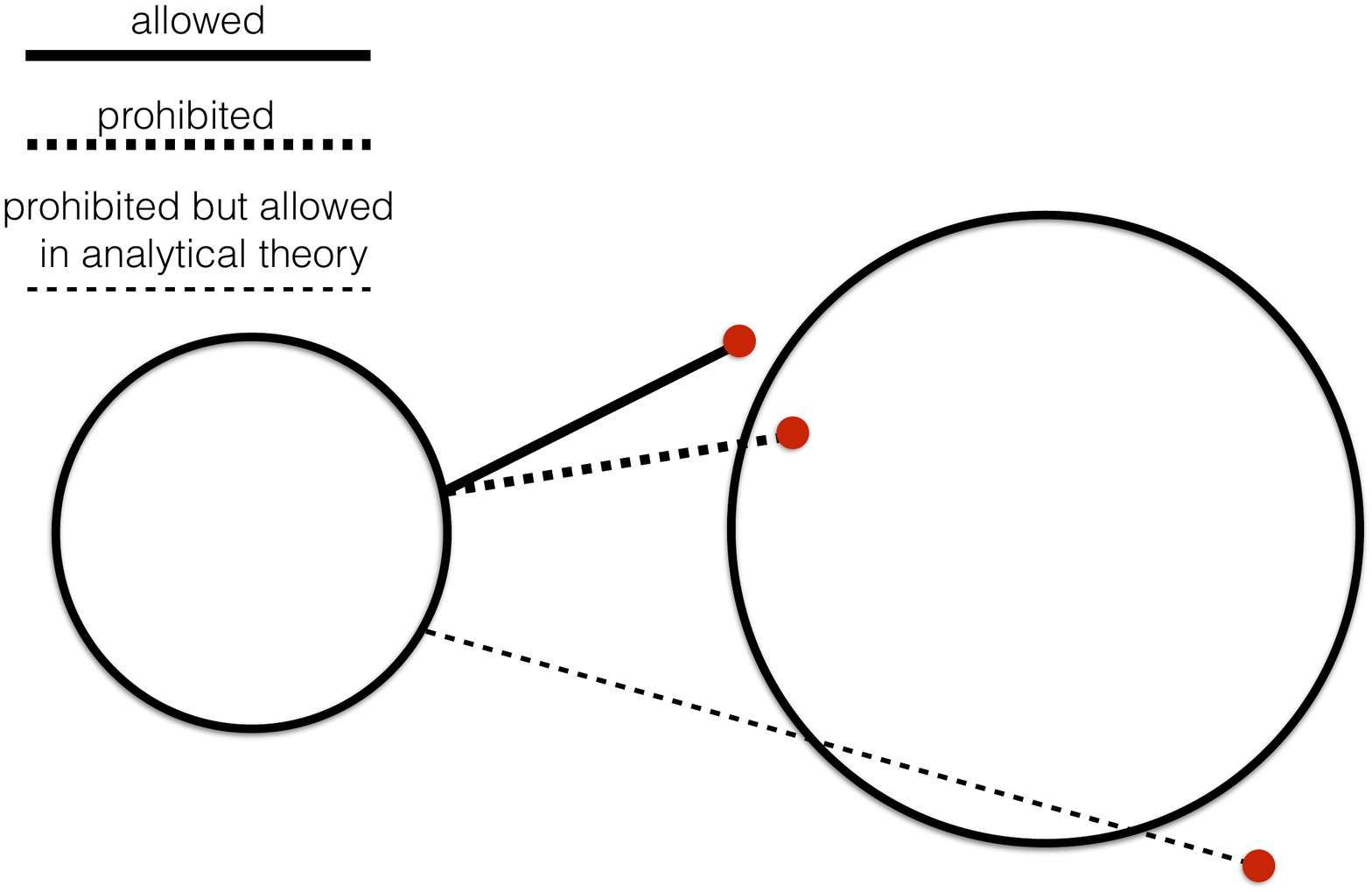} 
\caption{Example of a non-physical state of the polymer but included in the analytic calculation. 
The number of these states becomes smaller the smaller the ratio $\ell \over R_2$, where $R_2$ is the radius of the neighbouring colloid.}
\label{fig:1c}
\end{subfigure}

\caption{  }
\label{fig:approx}
\end{figure}

When colloids are functionalised with grafted polymers under good solvent conditions, polymeric chains 
act as steric stabilisers via excluded volume interactions. More precisely, they induce a repulsive free-energy between 
colloids due to the fact that the impenetrable colloids limit the amount of configurations the chains can attain, 
hence reducing the free-energy of the system. Formally, this can be written as:

\begin{equation}
\beta F_{rep} = -k_B T \ln\left( { \OR \over \Omega_{tot} } \right)
\label{eq:en_rep}
\end{equation}\\
where \OR is the partition function counting all accessible states of the 
polymers given the positions of all other colloids in the system $\{\vv{R_I}\}$ (see note \footnote{For non spherical colloids, their orientation should also be considered})
 and $\Omega_{tot}$ is \OR for isolated colloids (see Fig.~\ref{fig:1a} for reference). 
 We take the polymers to be ideal, and hence \OR simply counts the number of available geometric states, which are all considered to have exactly the same energy and hence
 the same weight, i.e. the system is athermal.
In our model, the polymer is a stiff, double-stranded (ds) DNA of length $\ell_i$ terminating with a small, point-like recognition sequence.
Since we assume $\ell_i<< \xi_i$ ($\xi_i$ being the persistence length of dsDNA), we can describe it as a rigid rod whose grafting point 
can move on the surface. As we are about to show, in this case a very accurate analytic approximation 
exists for Eq.~\ref{eq:en_rep}, which becomes better and better in the limit  $\ell/R \rightarrow 0$, $R$ being the radius of the colloid
on which the strand is grafted.\\

Let us first introduce a quantity, $P(\vv{r})$, defined as the probability that the end point of our rod 
(i.e that not grafted to the surface) is at a position \vv{r} (see Fig.~\ref{fig:1b} for reference).
$P(\vv{r})$ is given by the following expression:
\begin{equation}
P(\vv{r}) = \int_{-\pi}^{\pi} { 2 \pi R^2 \sin\left( \theta \right) d\theta \over 4 \pi R^2 } {   \delta\left( \mid \vv{r}-\vv{r}' \mid - \ell \right ) \over 2 \pi \ell^2 }  \Theta\left[ \left(\vv{r}-\vv{r}' \right) \cdot \left( \vv{r}'-\vv{O}\right)\right] 
\label{eq:pr}
\end{equation}\\
where the Dirac delta function takes care of fixing the rod length and the Heaviside step function makes 
sure that the rod does not penetrate the colloid on which is coated.
The integral in Eq.~\ref{eq:pr} can be calculated giving the following result:

\begin{equation}
P(\vv{r}) = { \begin{cases} {1 \over 4 \pi R \ell r } \qquad R^2 + \ell^2 < r^2 < ( R + \ell )^2; \\  0 \qquad \mathrm{otherwise.} \end{cases} }
\label{eq:pr2}
\end{equation}\\
which, for small $\ell / R \rightarrow 0$, can be further approximated as

\begin{equation}
P(\vv{r}) = { \begin{cases} {1 \over 4 \pi R^2 \ell } \qquad R^2 + \ell^2 < r^2 < ( R + \ell )^2; \\  0 \qquad \mathrm{otherwise.} \end{cases} }
\label{eq:pr3}
\end{equation}\\
i.e. $P(\vv{r})$ is \textit{uniform}. Our first approximation will be to take this uniform value for $P(\vv{r})$.
Let us now make a second approximation, whose validity also increases in the limit $\ell / R_2 \rightarrow 0$ (where 
$R_2$ now labels a colloid among the neighbours on which the polymer is grafted). 
We will count as allowed states for our rod all those for which its end-point is not inside a neighbouring colloid 
(see Fig.~\ref{fig:approx} for reference). Note that within this approximation we are (wrongly) counting as allowed 
configurations those where the rod overlaps a neighbour for a fraction which does not include 
the end point (see \ref{fig:1c}). 
Given that we take a flat probability distribution, the number of these states is simply proportional to the overlap 
volume $V_{overlap}$ between a sphere of radius $R_1 = R+\ell$ and that of a sphere of radius $R_2$, given by:

\begin{equation}
V_{overlap}(R_1,R_2,d) = {\pi \over 12 d } (R_1+R_2-d)^2 \left(d^2 + 2 d (R_1+R_2) - 3 (R_1^2+R_2^2) + 6 R_1 R_2 \right). 
\label{eq:voverlap}
\end{equation}\\
Since the total number of states available for a grafted rod when no other colloids are present is 
proportional to the volume accessible to its end-point $V_{tot} = 4 \pi R^2 \ell $, the following equation holds:

\begin{equation}
{ \Or \over \Omega_{tot} }  \approx V_{free} / V_{tot} = { V_{tot} - \sum_{neighbours} V_{overlap} \over V_{tot} } 
\label{eq:analytic1}
\end{equation}\\
hence, considering that rods behave as ideal (i.e. they do not interact with each other), we find the following final 
expression for the repulsive free-energy induced by $n_j$ mobile rods grafted on a surface of a colloid
due to the presence of its neighbours with positions $\vv{R}_{i}$:

\begin{align}
\beta F_{rep} &= -k_B T\ln\left(  \prod_{j=1}^{n_j}  { \Omega\left( \ell_j, \{\vv{R}_i \} \right) \over \Omega_{tot} } \right) \nonumber \\
		     &= -k_B T \sum\limits_j \ln\left( 1 - { \sum\limits_{i=1}^{N_{colloid}} V_{overlap}(R+\ell_j,R_i,d) ) \over 4 \pi R^2 \ell } \right).
\label{eq:analytic2}
\end{align}

\begin{figure}
\begin{center}
\includegraphics[width=\textwidth]{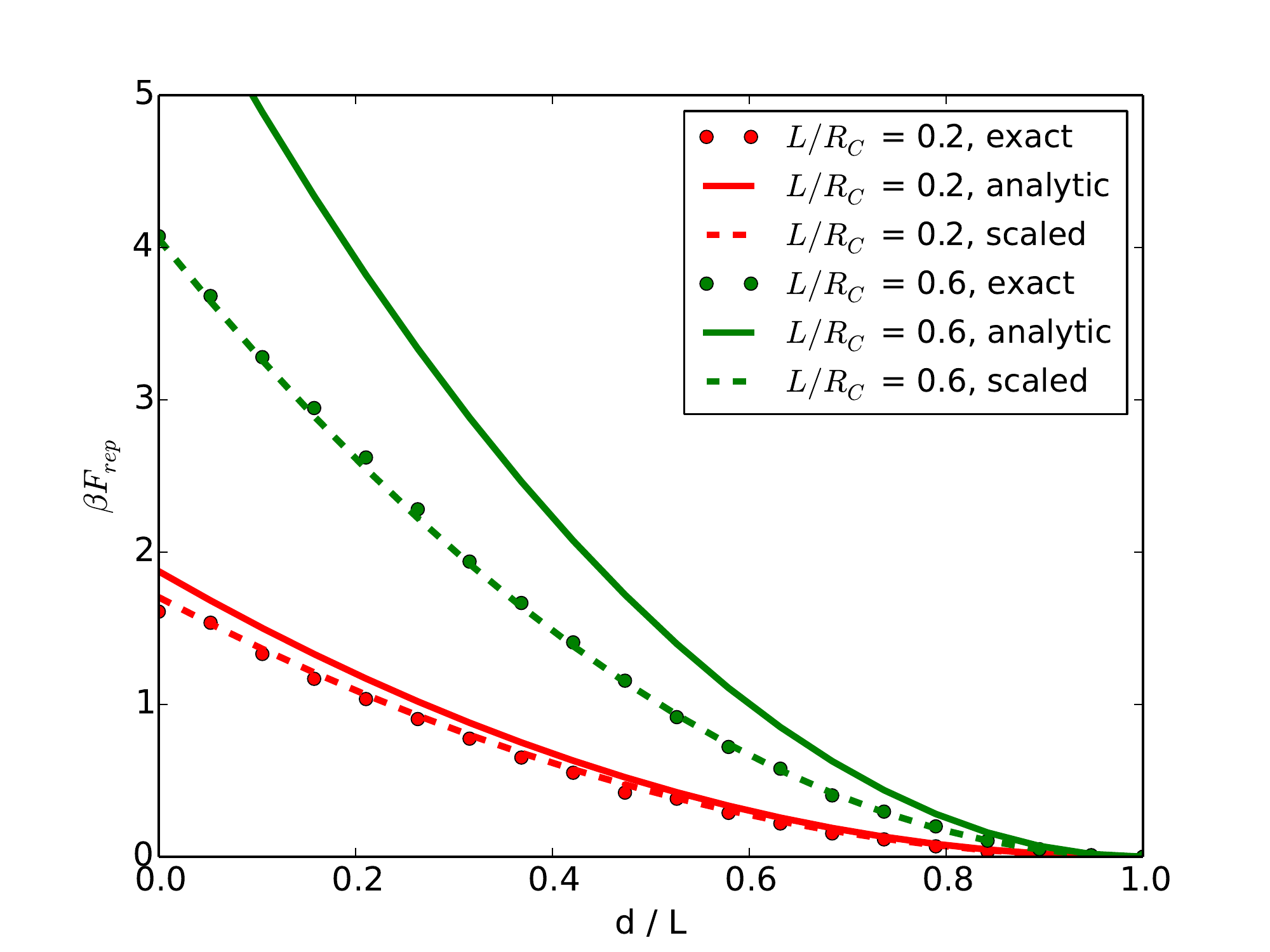} 
\end{center}
\caption{Comparison of the numerically accurate Monte Carlo evaluation of $F_{rep}$ as given by Eq.~\ref{eq:en_rep} (circles) 
and the analytic approximation given by Eq.~\ref{eq:analytic2}, with and without scaling to adjust it (dashed and full lines,respectively). Green
is for long strands ($L / R_C = 0.6$) whereas red is for the shorter ones ($L / R_C = 0.2$).
As expected, whereas the analytic approximation very well reproduces the Monte Carlo data for the shorter strands, a considerable difference appear
for the longer ones. However, a simple rescaling procedure is sufficient to reproduce the numerical estimation with high accuracy.
}
\label{fig:overlap3}
\end{figure}

To appreciate the difference between the analytic approximation and a numerically accurate Monte Carlo estimation of $F_{rep}$, 
we report in Fig.~\ref{fig:overlap3} the repulsive free-energy between two colloids calculated in both ways for two colloids coated
with 70 strands of either length $\ell=0.2~R_C$ or $\ell=0.6~R_C$, as in our simulations.

Whereas the difference is negligible for the short rods, it starts to become relevant for the longer ones. However, we observe that a simple
rescaling procedure can be used to get accurate results, i.e. one can take $F_{rep}^{scaled} = F_{rep}^{analytic} / \alpha $ and the data reproduces
the accurate numerical results for all colloids distances $d$ (for the longer strands $\alpha\approx 1.5$, note that its
value depends on the specific $\ell/R$ ratio).
This is equivalent to saying that, by using the analytical estimate for $F_{rep}$, we are effectively simulating a system whose grafting density is
about 1.5 times larger than the real one.
For the sake of computational efficiency and reproducibility, in our simulations we always use the analytic approximation. Hence, if
one intends to experimentally replicate the system, a higher coverage should be used (for the value of $\alpha$ in our model, this is still well in 
the range of values used in experiments).\\

As for $F_{rep}$, a similar approach can be taken in our system to calculate $\beta\Delta G^{cnf}$ (Eq.~4 in the main text), i.e. the
entropic configurational penalty to form a bond between two DNA strands. In our case, $\beta\Delta G^{cnf}$ is the configurational free-energy penalty 
to confine the end of two grafted rods at the same location (or, better, within a bonding volume $v_0$~\cite{bortolo-sm,patrick-jcp}). 
This is given by \cite{patrick-jcp}:

\begin{equation}
\beta\Delta G^{cnf} = - \log \left( {1 \over \rho_0 }{ \Omega_{ij} \over \Omega_i \Omega_j } \right)
\end{equation}\\
where $\rho_0$ is the standard concentration, $1$~M, $\Omega_{ij}$ is the phase space allowed for two strands $i$ and $j$ grafted at colloid $I$ and $J$ 
respectively and bound to each other, and $\Omega_{i(j)}$ the phase space allowed for two grafted but unbound strands, i.e. $V_{free}$ calculated previously. 
Using exactly arguments and approximations previously used to calculate $F_{rep}$, the following holds:

\begin{align}
\Omega_{ij} &= V_{overlap}( R_I+L_i, R_J+L_j , d) - V_{overlap}(R_I+L_i, R_J, d ) \nonumber \\ 
		   &- V_{overlap}(R_I, R_J+L_j, d )
\label{eq:omegaij}
\end{align}


\subsection{Large number of linkers limit of the bonding free-energy in the presence of mobile linkers:}

In this section we show that the set of self-consistent equations developed for DNA coated colloids (Eqs.\ 6 in the manuscript), and the free energy used (Eq.\ 7 in the manuscript)
are exact when the number of DNA-strands (linkers, from now on) becomes large.
This is shown by using a saddle--point approximation. 

\begin{figure}
\includegraphics[width=\textwidth]{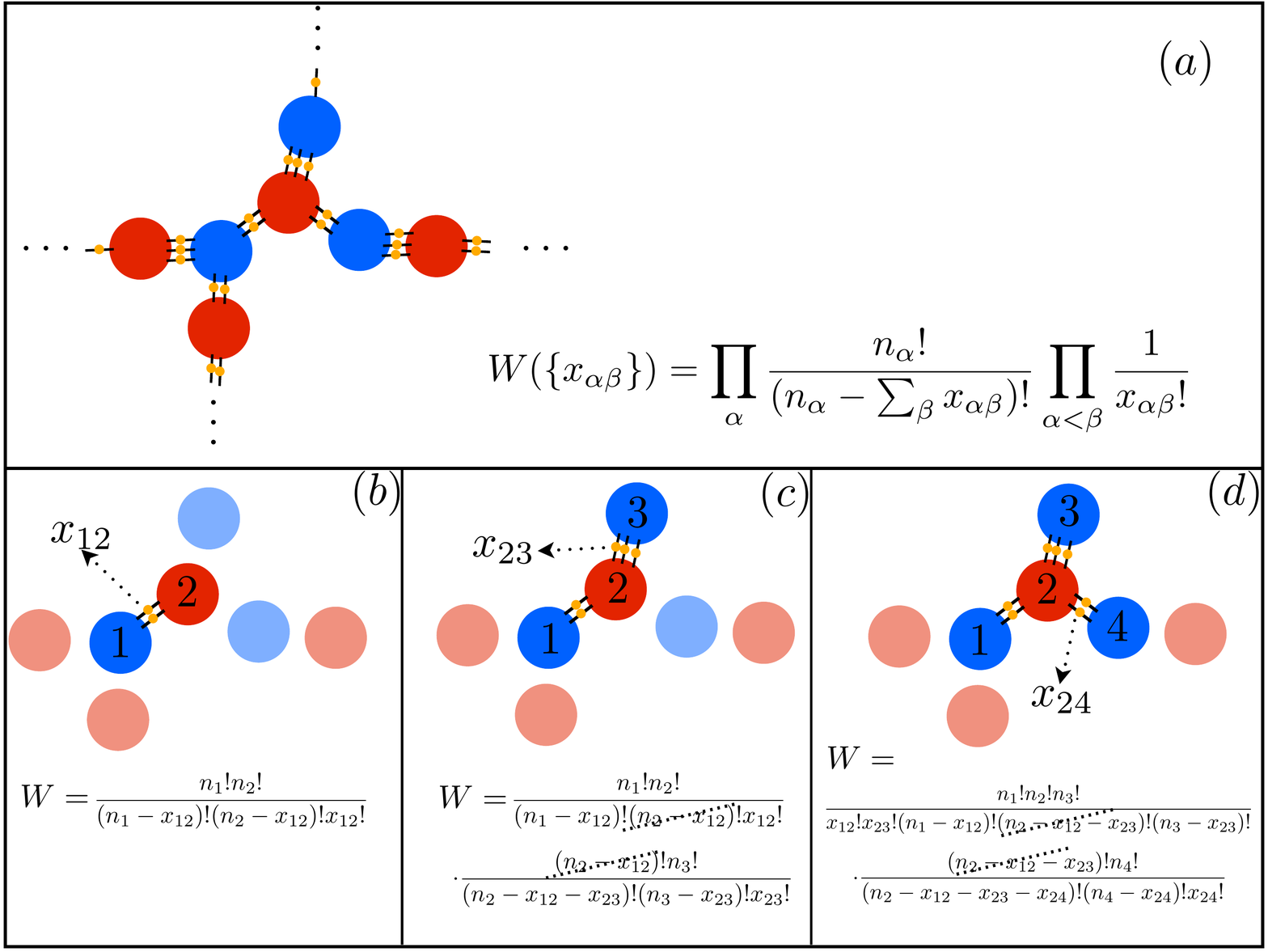} 
\caption{Counting the number of possible bonds combinations $W\left( \{x_{\aaa\bbb}\}\right)$ in Eq.~\ref{eq:z1} via combinatorial analysis (see procedure in the main text).}
\label{fig:SP}
\end{figure}

The DNA partition function for a set of colloids at fixed positions is given by 
\begin{eqnarray}
Z&=&\sum_{\{x_{\aaa\bbb}\}}  W(\{x_{\aaa\bbb}\}) (\Xi_{\aaa\beta})^{\sum x_{\aaa\bbb}} \, ,
\nonumber \\
W(\{x_{\aaa\bbb}\}) &=& \prod_\alpha {\na ! \over (\na-\sum\limits_\beta x_{\aaa\beta})!} \prod\limits_{\aaa<\beta} {1\over x_{\aaa\beta}!} \, ,
\label{eq:z1}
\end{eqnarray}
where  the $\Xi$ factors have been defined in Manuscript Eq.\ 5 (here we omit the dependence of $\Xi$ on the colloids' position).  $W (\{ x_{\alpha \beta}\})$ counts all the possible combinations of DNA--DNA hybridisation resulting in $x_{\alpha\beta}$ bridges between particle $\alpha$ and particle $\beta$ with $1\leq \alpha, \beta\leq N_\mathrm{type}$ (Fig.\ \ref{fig:SP}$a$). In the definition of $W$ (Eq.\ \ref{eq:z1}) the first product is taken on all the type of linkers while the second on all possible 
pairs of types. Linkers are of different types if they are either grafted on different colloids or have a different recognition sequences.
The partition function $Z$ is then defined as in Eq.~\ref{eq:z1} taking the sum on all the possible sets of pairs consistent with a total number of strands in the system weighted with the total hybridisation free energy (as given by Manuscript Eq. 5).

The derivation of $W(\{ x_{\alpha \beta}\})$ is shown in Fig.\ \ref{fig:SP}$(b-d)$ for the case in which different types correspond to different particles. The generalisation to the case in which more than one family of DNA is present on a single colloid is straightforward. 
Starting from a configuration with no bonds formed (in which $x_{\alpha\beta}=0$ for all $\alpha$ and $\beta$), the number of ways of making $x_{12}$ bridges between particle $1$ and particle $2$
is  given by (see Fig.\ \ref{fig:SP}$b$)  
\begin{eqnarray}
W_{12} = {n_1 ! n_2! \over (n_1 - x_{12})! (n_2 - x_{12})! x_{12} !} \, .
\nonumber
\end{eqnarray}
In the previous expression we have accounted for the  indistinguishability of the bridges
and $n_\alpha$ is the total number of linkers present on the particle $\alpha$.
At the next step we count the number of possible combinations resulting in  $x_{23}$ bridges between particle $2$ and $3$ finding
\begin{eqnarray}
W_{23} =  {(n_2-x_{12}) ! n_3! \over (n_3 - x_{23})! (n_2 - x_{12}-x_{23})! x_{23} !} \, . 
\nonumber
\end{eqnarray}
resulting in a total number of configurations (see Fig.\ \ref{fig:SP}$c$)
\begin{eqnarray}
W(x_{12},x_{23},0,\cdots,0) = W_1 W_2
=  {n_1! \over (n_1 - x_{12})!}{n_2!\over (n_2 - x_{12}-x_{23})!}{n_3!\over (n_3-x_{23})}{1\over x_{12} ! x_{23}!} \, .
\nonumber
\end{eqnarray}
In the previous expression we recognise the factorisation of Eq.\ \ref{eq:z1} in term of type terms (first three terms) and type--type terms (last two term). Recursively adding the missing bridges (e.g.\ see Fig.\ \ref{fig:SP}$d$ for $x_{24}$) it is easy to 
see that we obtain an expression for $W$ that remains consistent with Eq.\ \ref{eq:z1}.

Using a Stirling approximation we can write the partition function (Eq.~\ref{eq:z1}) as
\begin{eqnarray}
Z &=& \sum_{\{ x_{\aaa\bbb}\}} e^{-\beta {\cal F} (\{ x_{\aaa\bbb}\})}
\label{eq:ffun}
\\
\beta {\cal F} (\{ x_{\aaa\bbb}\}) &=& 
\sum_\aaa \Big[ \sum_\bbb x_{\aaa\bbb}+(\na -\sum_{\bbb} x_{\aaa\bbb}) \log (\na -\sum_{\bbb} x_{\aaa\bbb})
-\na \log \na 
\Big]
\nonumber \\
&&+\sum_{\aaa<\bbb}
\Big[
x_{\aaa\bbb} \log x_{\aaa\bbb} -x_{\aaa\bbb} -  x_{\aaa\bbb} \ln \Xi_{\aaa\bbb}
\Big].
\nonumber
\end{eqnarray}
In the limit in which $\na \to \infty$ (with $\na / \nb$ fixed) the number of linkers distribution functions peak around the saddle point solution $n_{\alpha \beta}$ defined by 
\begin{eqnarray}
{\partial 
\over \partial x_{\aaa\bbb}} {\cal F} (n_{\aaa\bbb}) = 0 \, .
\label{eq:sp1}
\end{eqnarray}
The free energy of the system ($\beta F$) is then given by 
\begin{eqnarray}
\beta F = \beta {\cal F}( \{ n_{\aaa\bbb} \} )
\label{eq:f1}
\end{eqnarray}

Using Eqs.~\ref{eq:sp1} and \ref{eq:ffun} we find
\begin{eqnarray}
n_{\aaa\bbb} = (\na-\sum\limits_\gamma n_{\alpha\gamma}) (\nb-\sum\limits_\gamma n_{\bbb\gamma})  \Xi_{\aaa\bbb} \, .
\label{eq:sp2}
\end{eqnarray}
Now we show that Eq.\ \ref{eq:sp2} is equivalent of Eq.~6 in the main text. This can be done 
identifying the probability that a tethers of type $\alpha$ to be free as 
\begin{eqnarray} 
p_\alpha = { n_\alpha - \sum\limits_\gamma n_{\alpha\gamma} \over n_\alpha } = {\overline n_\alpha \over n_\alpha} \, 
\nonumber
\end{eqnarray}
where $\overline n_\alpha$ is the number of tethers of type $\alpha$ unbound. Using  $p_\alpha$ we can write the tether balance as 
\begin{eqnarray}
\overline n_\alpha  + \sum_\beta n_{\alpha \beta} = n_\alpha
\nonumber \\
p_\alpha + \sum_\beta  {n_{\alpha \beta} \over n_\alpha} =1 \, .
\label{eq:tb}
\end{eqnarray}
Using Eq.\ \ref{eq:sp2} in Eq.\ \ref{eq:tb} we obtain the same set of equations found 
in the manuscript 
\begin{eqnarray}
p_\alpha + \sum_\beta n_\beta p_\alpha p_\beta \Xi_{\alpha \beta} =1 \,
\end{eqnarray}
with $\alpha=1,\cdots, N_\mathrm{type}$.

We are now in a position  to evaluate the free energy \ref{eq:f1}. 
Using Eq.\ \ref{eq:sp2} we can write
\begin{eqnarray}
\sum_{\aaa<\bbb}  \Big[ \naabb \log \naabb 
-\naabb \log \Xi_{\aaa\bbb}  \Big]
 &=& \sum_{\aaa<\bbb} n_{\aaa\bbb}\Big[\log (\overline n_\aaa \overline n_\bbb \Xi_{\aaa\bbb}) - \log \Xi_{\aaa\bbb}  \Big]
\\
&=& \sum_{\aaa<\bbb} n_{\aaa\bbb}\Big[\log \overline n_\aaa  + \log n_\bbb  \Big]
\\
&=& \sum_{\aaa,\bbb} n_{\aaa\bbb} \log \overline n_\aaa 
\end{eqnarray}

Using the previous equation in \ref{eq:ffun} we find
\begin{align}
\beta F &= {1 \over 2} \sum_{\aaa,\bbb} \naabb  + \sum_{\aaa} n_\aaa \log{\overline n_\aaa \over n_\aaa}  \nonumber\\
\beta F &= {1 \over 2 } \sum_{\aaa} \Big( n_\aaa - \bna  + n_\aaa \log p_\aaa \Big) \nonumber\\
\beta F &= {1 \over 2 } \sum_{\aaa} n_\aaa (1 - p_\aaa) + \sum_\aaa n_{\aaa} \log p_\aaa \nonumber \\
\beta F &= \sum_{\aaa} n_\aaa \left[ \log(p_\aaa) + 1/2 \left(1 - p_\aaa\right)\right] 
\label{eq:free-energy-final}
\end{align}

which again  is equivalent to Eq.~7 in the main text.

In Ref.~\cite{stefano-jcp}, Equations~1-3 of the main text, from which the final self-consistent equations and the free-energy of the system of mobile 
linkers are obtained here, were derived assuming that the conditional probability for a linker to be unbound given that another one was unbound was independent 
from the latter. Since for the case of mobile linkers we obtain the same functional equations considering an \textit{exact} partition function, this means that in this
latter case the approximation becomes  \textit{exact} in the limit where a large number of linkers is present.

\subsection{An analytical expression for $F_{clus}$}

Eq.~8 in the main text can be solved to give:

\begin{align}
\begin{cases}
p_\aaa &= { -1 + \Xi ( n_\aaa - N_B n_\bbb ) + \sqrt{4 \Xi n_\aaa + \left (1-\Xi \left( n_\aaa - N_B n_\bbb \right)\right)^2  }
\over 2\Xi n_\aaa } \\ 
p_\bbb &= { -1 + \Xi ( N_B n_\bbb - n_\aaa ) + \sqrt{4 \Xi n_\aaa + \left (1-\Xi \left( n_\aaa - N_B n_\bbb \right)\right)^2  }
\over 2 \Xi N_B n_\bbb } 
\end{cases}
\label{eq:sol_pi}
\end{align}

Combining this solution to Eq.~9 in the main text, the following formula arises for the bonding part of the free-energy of a cluster of an $A$ colloid coated with $n_\aaa$ linkers
surrounded by $N_B$ neighbours, each coated with $n_\bbb$ linkers, in equivalent positions (linkers~\aaa~and~\bbb~are complementary and can form a
bond with average energy $\Xi$, see main text for details). 

\begin{align}
F_{clus}  &= n_\aaa (A_1 + A_2 ) + N_B n_\bbb (B_1 + B_2 ) + N_B F_{rep}^{min}\\
A_1 &= 1/2 \left( 1 - { \Xi^{-1} ( -1 + \Xi n_\aaa - \Xi N_B n_\bbb  + \sqrt{4 \Xi n_\aaa + ( 1 + \Xi ( -n_\aaa + N_B n_\bbb) )^2}  )   \over 2 n_\aaa  } \right) \\
A_2 &= \log\left( { \Xi^{-1} ( -1 + \Xi n_\aaa - \Xi N_B n_\bbb  + \sqrt{4 \Xi n_\aaa + ( 1 + \Xi ( -n_\aaa + N_B n_\bbb) )^2}  )   \over 2 n_\aaa  } \right)   \\
B_1 &= 1/2 ( 1 - { \Xi^{-1} ( -1 - \Xi n_\aaa + \Xi N_B n_\bbb  + \sqrt{4 \Xi n_\aaa + ( 1 + \Xi ( -n_\aaa + N_B n_\bbb) )^2}  )   \over 2 N_B n_\bbb  } \\
B_2 &= \log\left( { \Xi^{-1} ( -1 - \Xi n_\aaa + \Xi N_B n_\bbb  + \sqrt{4 \Xi n_\aaa + ( 1 + \Xi ( -n_\aaa + N_B n_\bbb) )^2}  )   \over 2 N_B n_\bbb  } \right) 
\end{align}


\bibliography{biblio-mobile}

\end{document}